# Inelastic collision-induced atomic cooling and gain linewidth suppression in He-Ne lasers

Yuanhao Mao[a,b,†], Jipeng Xu[a,b,†], Shiyu Guan[a,b], Hongteng Ji[a,b], Wei Liu[a,b,*], Dingbo Chen[a,b], Qiucheng Gong[a,b], Yuchuan Quan[c], Xingwu Long[a,b], Hui Luo[a,b,*], Zhongqi Tan[a,b,*]

[a]College of Advanced Interdisciplinary Studies, National University of Defense Technology, Changsha, Hunan 410073, China.

[b]Nanhu Laser Laboratory, National University of Defense Technology, Changsha 410073, China.

[c]Huaguan Technology Co, Ltd., Changsha 410073, China.

[†]These authors contributed equally to this work.

[*]Email address: wei.liu.pku@gmail.com, luohui.luo@163.com, zhqitan@sina.com

**Abstract.** He-Ne lasers have been one of the most widely employed optoelectronic elements, playing irreplaceable roles in various applications, including optical detections, spectroscopy, interferometry, laser processing, and so on. For broad applications that require single-mode operations, the gain linewidth needs to be constrained, which conventionally can be obtained through overall gain suppressions. Such an approach inevitably has limited the output power and thus restricted further applications that require ultra-high precisions. In this article, we discover that inelastic collisions among He and Ne atoms can be exploited to cool down the Ne atoms, compressing the Doppler broadening and consequently also the gain linewidth, enabling us to further experimentally demonstrate a significantly broadened spectral range of single-mode operation with stable output powers. Our discovery of inelastic collision-induced atomic cooling has ultimately overcome the tradeoff between output power and gain linewidth, opening new avenues for both fundamental explorations and disruptive applications relying on gaseous laser systems.

## 1 Introduction

As the world's first gas laser ever invented, the He-Ne laser has been playing a vital role in high-precision measurements and sensing, including the establishment of length and time-frequency standards [1–5], large-scale interferometric measurements [6–9], atomic and molecular spectroscopy [10–13], gravity measurements [14–16] and inertial navigation [17–19]. A prominent application of the He-Ne laser is the large ring laser gyroscopes (RLGs), which are widely employed for rotational seismic wave monitoring [20–23], universal time (UT1) measurement [9,24,25] and gravitational wave detection [26–28].

To obtain ultrahigh-precision measurement, larger laser interferometers (including but not limited to RLGs) are usually required. Laser cavities of larger sizes could indeed enhance detection sensitivity, but at the same time also lead to smaller inter-mode intervals and thus multi-mode competitions [29]. This would inevitably reduce the coherence length [30] and increase the output power instability [31,32]. With the cavity size fixed, the conventional approach to obtain a larger spectral range of single longitudinal mode (SLM) oscillation is to reduce the pump intensity, which effectively narrows the gain linewidth (see the left column of Fig. 1). However, this approach of gain suppression would sacrifice the output power, leading to amplified quantum noises and reduced signal-to-noise ratio [17,33]. Such a profound tradeoff between the output power and gain linewidth poses the central obstacle for further applications relying on He-Ne and more generally gaseous lasers.

In this article, we discover a new mechanism based on atomic cooling to overcome this fundamental tradeoff. It is shown that inelastic collisions and resonance energy transfer (RET) among He and Ne atoms [34] can effectively cool down the Ne atoms, compressing the Doppler-broadened gain linewidth. An enlarged spectral range of SLM oscillations is experimentally observed, without sacrificing the output power. Theoretical analysis has further verified the mechanism of inelastic collision-induced atomic cooling underlying our experimental observations. Our results can trigger further fundamental explorations into similar lasing dynamics based on atomic collisions, and spawn novel and disruptive applications relying on gaseous laser systems.

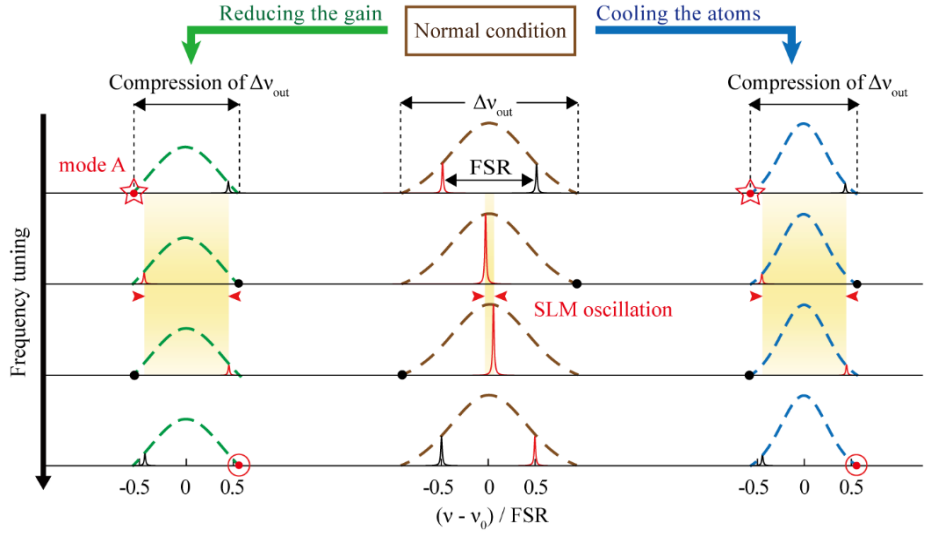

**Fig. 1. Different approaches [reducing the gain (left column) and cooling the atoms (right column)] to expand the spectral ranges of SLM oscillations (light yellow shaded areas).** The dashed lines denote the gain curves, and $\Delta \nu_{out}$ is the corresponding output spectral width; the pentagrams and the circles mark respectively the birth and death points of lasing mode A (represented by the red curves), viewed from lower to higher frequencies; black dots and curves denote the resonant positions of the neighbouring modes (represented by the black curves); FSR (free spectral range) characterizes the spectral distance between adjacent longitudinal modes.

## 2 Introduction

### 2.1 Approaches to expand the spectral ranges of SLM oscillations

For a laser system, the spectral region (denoted by the dashed gain curves in Fig. 1) where the modes can be in the lasing states is characterized by the output spectral width $\Delta \nu_{out}$. As is shown in Fig. 1, the boundaries of this region correspond to the birth and death positions (denoted by pentagrams and the circles, respectively; viewed from lower to higher frequencies) of the lasing modes (denoted by the red curves). The spectral region of SLM oscillations is marked by the shaded areas (light yellow), where only one mode is in the lasing state (red curves): when the lasing mode is located on the boundaries of this region, there is a neighbouring mode (denoted by the black dots) that just enters or leaves its own lasing region. The spectral spacing between two neighbouring modes is termed as free spectral range (FSR), as is shown in the middle column (top) of Fig. 1.

One approach to compressing $\Delta v_{out}$ and thus expanding SLM regions is to reduce the gain strength of the lasing system (see left column of Fig. 1: shifting the gain curves downward without compressing itself). The obvious sacrifice of such a method is the reduced output power. An alternative approach to compress $\Delta v_{out}$, with the output power maintained, is to suppress the gain linewidth broadening through atomic cooling (see right column of Fig. 1: compressing the gain curves without shifting downward).

**Measurement of gain linewidth**
Temperature is essentially a measure of the average speed of atomic or molecular motions. In a gas gain medium, the linewidth of the Doppler broadening $\Delta v_D$ is directly related to the average velocity $\bar{v}$ of atoms [30]:

$$\Delta v_D = v_0 \sqrt{ln2 \frac{8k_b T}{mc^2}} = \frac{\bar{v}}{\lambda} \sqrt{\pi ln2} \tag{1}$$

where $k_b$ is Boltzmann constant; $T$ is atomic temperature; $m$ is atomic mass; $c$ is the speed of light; $v_0$ and $\lambda$ is the central frequency and wavelength of the Doppler broadening line. For the whole laser system, the total gain linewidth is the convolution of the Doppler broadening and the homogeneous broadening [35]. As a result, with increasing gas pressures and thus also enhanced homogeneous broadening, compression in the total gain linewidth can be only induced by suppressions of the Doppler broadening. That is to say, under the conditions of higher gas pressures, the observation of narrower gain linewidth would be a direct indicator for atomic cooling according to Eq. (1).

The overall gain curve of a laser can be measured through frequency tuning. As has been already shown in Fig. 1, the spectral width between the birth and the death points of a specific lasing mode corresponds to $\Delta v_{out}$ (also shown in Fig. 2b) (**More details of the measurement principle are included in Supplemental Materials S2.**). At the same time, the intensity variation of a single lasing mode with tuning frequencies directly corresponds to the gain curves of the laser.

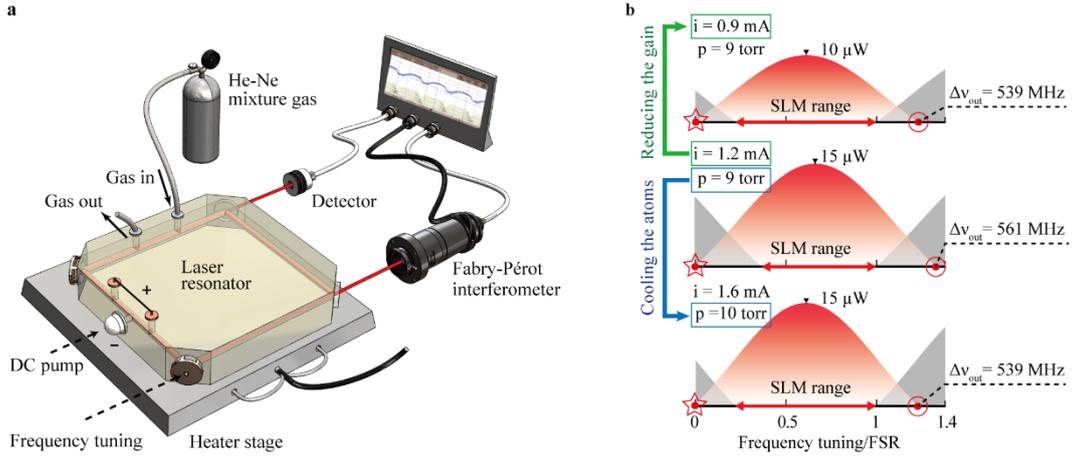

**Fig. 2. Experimental setup and methods of compressing the output spectral widths $\Delta\nu_{out}$. (a)** A square laser cavity is fixed on a heater stage with inlet/outlet gas ports. Spectral and intensity signals are obtained by a Fabry-Pérot interferometer and a photodetector. **(b)** The output spectral width $\Delta\nu_{out}$ can be compressed by reducing the pump current with the pressure fixed (indicated by the green arrow: reducing the gain) or injecting more He-Ne mixed gas and increasing the pump current (indicated by the blue arrow: cooling the atoms).

Figure 2a shows the experimental setup, in which the laser resonator is a square cavity with an FSR = 417 MHz. The cavity is sealed with He-Ne gas and has inlet/outlet ports to control the total pressure and the ratio between He and Ne atoms. The laser is pumped by a DC power, enabling precise manipulation of the pump current to adjust the gain. Frequency tuning is achieved using a pair of piezoelectric ceramics connected to the two spherical mirrors on the right side of the cavity. A Fabry-Pérot interferometer and a photodetector are used for spectral and intensity signal acquisition through two output mirrors on the left side. As is shown in Fig. 2b, as the pump current drops from 1.2 mA to 0.9 mA with the pressure fixed (indicated by the green arrow in Fig. 2b): the output spectral width decreases from 561 MHz to 539 MHz, and the peak output power has also been reduced from 15 μW to 10 μW. In contrast, the same compression of $\Delta\nu_{out}$ can be achieved by increasing both the pump intensity (from 1.2 mA to 1.6 mA) and cavity pressure (from 9 torr to 10 torr), without sacrificing the peak output power which is maintained at 15 μW (indicated by the blue arrow in Fig. 2b).

In the SLM regime, when the central lasing frequency coincides with the central position of the Ne atomic gain curve, the output intensity reaches its maximum value. This maximum intensity reflects the strength of the maximum pure gain. When the maximum gain is stable (with constant maximum output power), the broadening or narrowing of output spectral widths $\Delta\nu_{out}$ directly correspond to the same variation of overall gain linewidth. As has been already clarified, with higher pressure and thus enhanced homogeneous broadening, the narrowing of $\Delta\nu_{out}$ could only originate from atomic cooling-induced Doppler broadening suppression: this is exactly what has been observed in Fig. 2b, as indicated by the blue arrow. This is to say, injecting more He-Ne mixed atoms in a confined space (with higher pressure) and at the same time increasing pump currents (without increasing the pump current the peak output power would not be maintained; **see Supplemental materials S3**) represent a new scheme for atomic cooling.

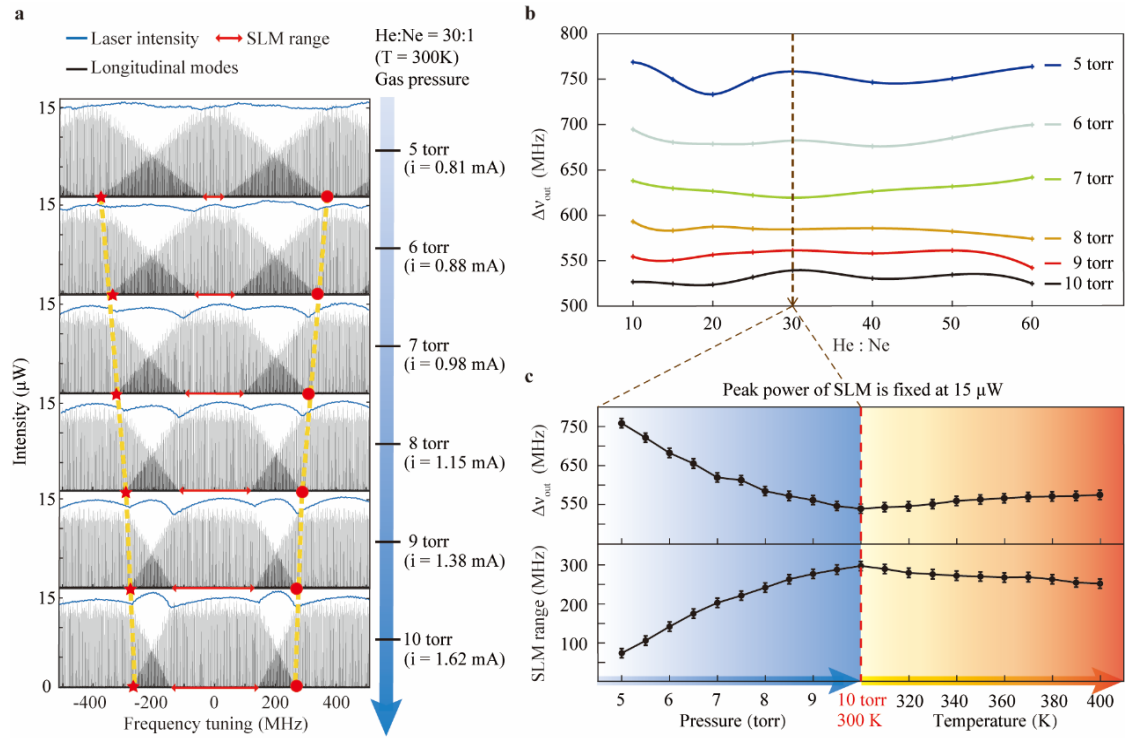

**Fig. 3. Experimental results with different pressures, gas ratios and varying temperatures.** (a) The spectral measurement shows that the output spectral width is reduced (indicated by yellow dashed lines), while the SLM oscillation region is broadened with an increase of pressure (injecting more gas) and pump current for a mixed He-Ne gas (He:Ne = 30:1). The temperature is maintained at 300 K. (b) The dependence of $\Delta\nu_{out}$ on different gas ratios under different pressures, where the peak output power is maintained at 15 μW. c The evolution of output spectral width and SLM range with increasing pressure (temperature is constant) and increasing temperature (total gas atom number is constant).

The detailed spectral evolutions of the He-Ne laser with increasing pressure and pump current are shown in Fig. 3a, with the peak power of SLM maintained at 15 μW, He-Ne ratio fixed at 30:1, and a constant temperature of 300 K. It is evident that as the pressure increases, the output spectral width of the laser continuously decreases, while the SLM range continuously increases. Such a mechanism can be employed to significantly improve the stability of optical sensor devices such as RLGs through suppressing modes competition. In addition, compared to lower pressures (6~8 torr), when the pressure exceeds 9 torr, the effect of narrowing $\Delta\nu_{out}$ almost saturates and is no longer significant anymore.

This can be attributed to enhanced homogeneous broadening caused by more and more atomic collisions, which essentially has offset the effect of suppressed Doppler broadening. That is, increasing the pressure to compress the gain linewidth will become less effective when homogeneous broadening becomes dominant.

Figure 3b further shows the dependence of gain linewidth on He-Ne gas ratios. It is clear that compared to the effect of increasing pressures, the variations of $\Delta\nu_{out}$ induced by change of gas ratio is minor.

We have so far confined our discussions to a constant temperature of 300 K and increasing pressure is obtained by injecting more He and Ne gas into the cavity. As a next step, we maintain the gas amount within the cavity and investigate the effect of changing temperature. As is shown in Fig. 3c (right half), once the pressure reaches 10 torr with a constant temperature of 300 K, the laser resonator is sealed to maintain a constant number of He-Ne atoms. The resonator is then gradually heated, and it is observed that the output spectral width and SLM range exhibit opposite trends (increase and decrease, respectively). The peak output power is maintained at 15 μW throughout the whole process. That is to say, with a fixed atom number and peak output power, reducing the temperature to cool the atoms also narrows the output spectral width. These temperature variation experiments provide further evidence that increasing pressure without changing the temperature indeed results in inelastic collision-induced atomic cooling, with an effect similar to that of reducing the temperature.

**Simulation of resonance energy transfer (RET) reactions**

The gain of He-Ne lasers originates from a typical physical process that involves inelastic collisions between atoms, where Ne atoms are pumped to the excited state through the resonance energy transfer (RET) reaction [30,34]:

$$He^*(1s2s) + Ne \rightarrow Ne^*(2s^2 2p^5 5s) + He - \Delta E(0.047 \text{eV}) \qquad (2)$$

where * refers to the excited states of atoms. To achieve population inversion, the RET reactions require the conversion of part of the kinetic energy to the atomic potential energy to account for the energy difference between He* and Ne*. **(The detailed energy level of He-Ne producing 632.8 nm laser can be found in Supplemental Materials S1.)** Subsequently, Ne* atoms release energy through stimulated or spontaneous emission and eventually return to the ground state through wall collisions [30,34]. When the laser output power is stable, the velocity distribution of ground state Ne atoms will reach a dynamic equilibrium, which is different from that of an ideal adiabatic system. That is, the system converts part of its kinetic energy to lasing photons, which leads to atomic cooling and thus broadened SLM region.

Our previous experimental results have demonstrated that Ne atoms can be cooled by injecting more He-Ne mixed atoms into the resonator. We now try to verify theoretically that this atomic cooling stems from the kinetic energy consumption during the RET reactions, causing more Ne atoms to be cooled in a local space filled with Ne-He atoms under higher-pressure conditions. We have employed Monte Carlo methods to simulate the inelastic collision process described by Eq. (2) to demonstrate that higher gas pressure (more atoms in a confined space) leads to more intense inelastic collisions, reducing the average velocity of Ne atoms in the steady-state equilibrium.

To quantitatively analyze the effect of RET reactions on Ne atoms, the simulation is confined within a 1 μm³ cubit unit cell (Fig. 4a) to track the whole particle motions over a time range of 0~200 ns with a time resolution of 1 ps. In order to effectively converge the simulations, some simplifications are implemented: (i) Firstly, considering that the number of He atoms is much greater than that of Ne atoms, it is assumed that Ne atoms always collide with He* that is excited by electrons; (ii) In the meantime, Ne* atoms are set to return to the ground state after 100 ps; (iii) A periodic boundary condition is adopted due to the spatial translation symmetry of atom distributions.

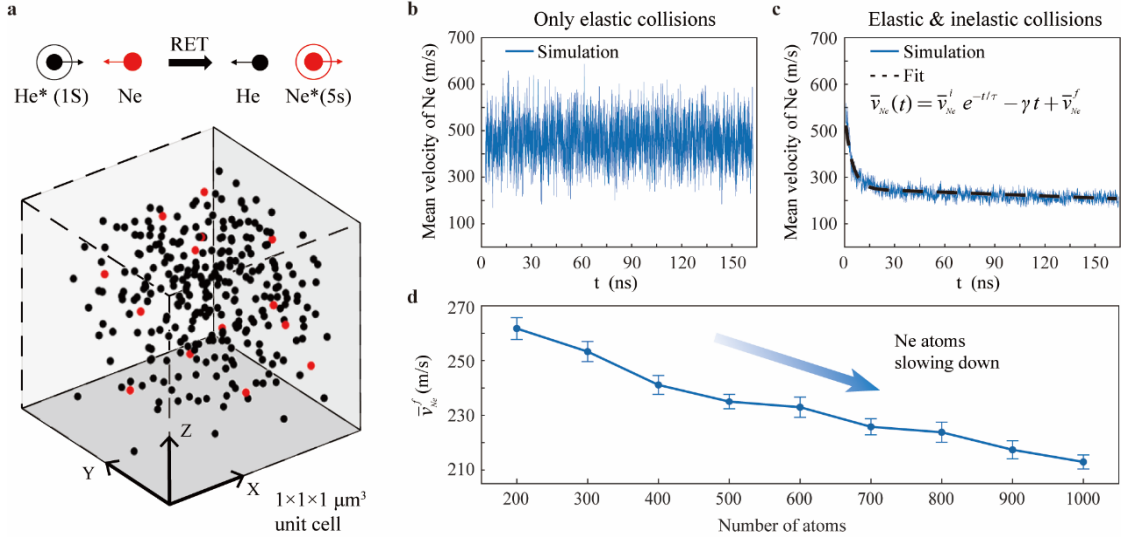

**Fig. 4. Monte Carlo simulations of RET reactions between He\* and Ne. (a)** Initial 1 μm³ cubic unit cells containing a large number of He and Ne atoms; **(b), (c)** The average velocity of Ne atoms ($\bar{v}_{Ne}$) is recorded every picosecond with only elastic collisions (in **b**) and elastic combined with inelastic collisions (in **c**); **(d)** Ne atoms are slowing down with more atoms in a confined space due to higher frequency of inelastic collisions.

In this simulation, we continuously track all atomic collisions, including the total number of collisions and inelastic collisions specifically between He\* and Ne. The average velocity of Ne atoms, $\bar{v}_{Ne}$, is also recorded. As shown in Fig. 4b, c, in contrast to the atomic motions with only elastic collisions, when there is a RET process among atoms, $\bar{v}_{Ne}$ drops significantly before reaching the equilibrium. An exponential function ($\bar{v}_{Ne}(t) = \bar{v}_{Ne}^i e^{-t/\tau} - \gamma t + \bar{v}_{Ne}^f$, where $\bar{v}_{Ne}^i$ is the initial average velocity of Ne atoms; $\tau$ is the relaxation time; $\gamma$ is a coefficient related to energy dissipation) is used to fit the evolution process of Ne atoms, with both inelastic and elastic collisions taken into consideration. Then the final average velocity of Ne atoms ($\bar{v}_{Ne}^f$) in the equilibrium can be obtained. Fig. 4d demonstrates that in a confined space, the increasing number of atoms will result in a decreasing trend of $\bar{v}_{Ne}^f$, where the error bar indicates the confidence interval for the fitting parameters of the simulation results. All in all, our Monte Carlo simulations have verified that higher gas pressure can lead to a decrease in the average velocity of Ne atoms. Consequently, an effective compression of Doppler broadening and expansion of the SLM region has been achieved.

**Conclusion**

In summary, we have demonstrated for the first time that in He-Ne lasers, the RET process involving inelastic collisions among atoms can be employed to cool down the Ne atoms. Such atomic cooling can effectively reduce the gain linewidth and consequently broaden the spectral regions of SLM oscillations. Our discovered mechanism can be further exploited to improve the functionalities of many other gaseous lasing systems, such as enhancing the stability and signal-to-noise ratio of large ring laser gyros that have been playing a crucial role in gravitational wave detection and seismic wave monitoring. Our results can surely incubate both fundamental explorations and disruptive applications in the general fields of laser physics and technologies.

**Materials and methods**

## Details of the experimental setup

The main body of the laser resonator shown in Fig. 2a is manufactured from Zerodur, a glass ceramic with a thermal expansion coefficient <10$^{-8}$/K. To characterize the changes in the gain curve as accurately as possible, four reflective mirrors are super polished to sub-nanometer roughness [36], whose total loss is less than 300 ppm. The FSR of the Fabry-Pérot interferometer is 1.5 GHz and its fineness is above 250. In order to maintain the uniformity of heating, we designed a thermo-insulation box to heat the laser, and the temperature distribution inside the box is simulated in **Supplemental Materials S5**. Additionally, whenever the temperature gradually increases to a test point, we will keep warm for 30 minutes before performing spectral measurements.

## Measurements of gain curves

The Fabry-Pérot interferometer is driven by a 10 Hz triangular wave signal, while the tuning frequency of the laser resonator is 0.2 Hz, which allows continuous recording of the output spectrum at different moments. When there is only one longitudinal mode oscillating, the output power reflects the pure gain. Therefore, as long as we arrange the evolution of a particular longitudinal mode with the tuning frequency in the same coordinates, the contour formed by the vertices of the longitudinal mode is the gain curve. Details of spectral measurements can be found in **Supplemental Materials S2**.

## Monte Carlo simulation

The simulation is based on the equation of motion of the inelastic collision between He* and Ne atoms:

$$\begin{cases} m_{He} v^i_{He} + m_{Ne} v^i_{Ne} = m_{He} v^f_{He} + m_{Ne} v^f_{Ne} \\ \frac{1}{2} m_{He}(v^i_{He})^2 + \frac{1}{2} m_{Ne}(v^i_{Ne})^2 = \frac{1}{2} m_{He}(v^f_{He})^2 + \frac{1}{2} m_{Ne}(v^f_{Ne})^2 - \Delta E \end{cases}$$

Only when the He* and Ne atoms have considerable relative motion velocities can the RET reaction be triggered: $(v^i_{He} - v^i_{Ne})^2 \geq \frac{2(m_{He}+m_{Ne})}{m_{He} m_{Ne}} \Delta E$, otherwise, only elastic collisions will occur. The initial velocities of He* and Ne atoms are set randomly according to the Maxwell-Boltzmann distribution, with average velocities of 989.9 m/s and 422.1 m/s respectively, corresponding to the temperature of 300 K.

## Acknowledgements

Z.T. acknowledges the support from Science Foundation for Indigenous Innovation of National University of Defense Technology (22-ZZCX-063), Natural Science Foundation of Hunan Province of China (2023JJ30639) and Natural Science Foundation of China (62375285). Y.M. acknowledges Mr. Yunjin Guan and Mr. Hui Hu for their full supports on experimental setup.

## Author details

[1]College of Advanced Interdisciplinary Studies, National University of Defense Technology, Changsha, Hunan 410073, China.
[2]Nanhu Laser Laboratory, National University of Defense Technology, Changsha 410073, China.
[3]Huaguan Technology Co, Ltd., Changsha 410073, China.

## Author contributions

Y.M., J.X., W.L., Z.T. and H.L. proposed and designed the research. Z.T., Y.Q., X.L. and Q.G. designed and fabricated the high-quality He-Ne laser resonator. Y.M., J.X., S.G., D.C. and H.J. did the spectral measurements and analyzed the raw data. Y.M., J.X. did the theoretical analysis and simulations. Y.M.,

J.X. and W.L. wrote the manuscript. All authors discussed the results and commented on the manuscript.

**Conflict of interest**

We are preparing for a patent application based on the work described here.

## Supplementary materials

### S1: Energy level of He-Ne laser

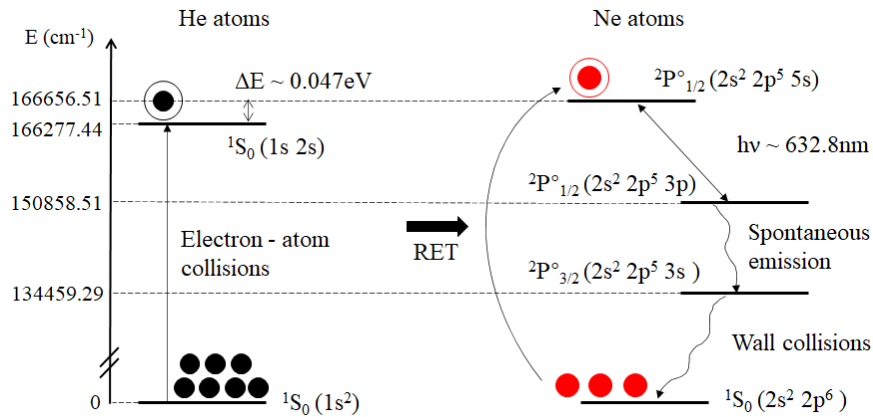

Fig. S1 **Schematic diagram of the He-Ne laser (@632.8nm) generation mechanism.** RET: Resonance energy transfer.

As is shown in Fig. S1, the He-Ne laser is a typical four-level laser system, taking the 632.8 nm laser generation process as an example: firstly, electrons are accelerated by DC or AC power supply to pump He atoms from the ground state to the excitation state He* (1s2s, 20.9 eV), whose energy level is very close to that of Ne* ($2s^2 2p^5 5s$), with about 0.047 eV energy difference. So the RET process will occur with a large scattering section between He* and Ne atoms at the expense of kinetic energy.

Inelastic collisions between He* and Ne trigger this RET reaction, which needs kinetic energy to compensate for the difference in energy level between He* and Ne*. This step is the main source of pumping Ne atoms from the ground state to the upper energy level, which constitutes a prerequisite for laser generation. Atoms at the lower energy level can only return to the metastable state ($2s^2 2p^5 3p$) through spontaneous radiation, and this metastable state requires an energy dissipation process through wall collisions to finally return to the ground state.

### S2: The method of measuring the gain curve of a laser

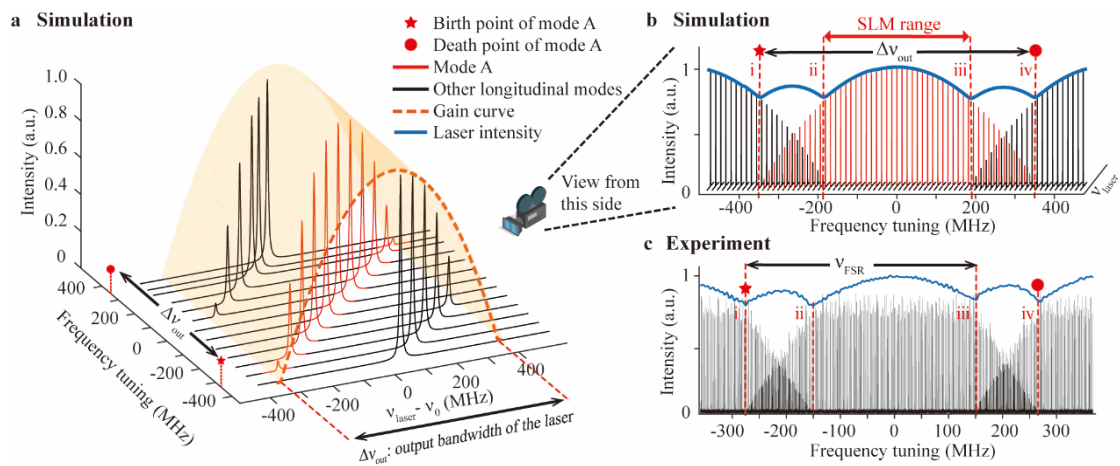

Fig. S2 **Method of measuring the gain curve. a,b** Simulation of laser modes evolution by tuning a resonator under a certain gain (a) and a view from the side of frequency tuning (b). **c** An experimental result of longitudinal modes and laser intensity evolving with tuning frequency under the condition of mixed gas (He:Ne = 10:1) with 9 torr pressure. SLM: single longitudinal mode.

The measurement principle is shown in Fig. S2a, where we performed a simulation of the tunable spectrum of a long-cavity laser capable of exciting dual longitudinal modes, with $v_0$ being the centre of the gain curve. To generate laser emission, the condition of a positive gain-loss difference must be satisfied. Therefore, the intensity variation of a single laser mode (red curves in Fig. S2a, b) with tuning frequency represents the gain curve of the laser (shown as the orange surface in Fig. S2a). The tuning range of the resonator from the birth point (pentagram) to the death point (circle) of a single laser mode corresponds to the output bandwidth of a laser, $\Delta v_{out}$.

Fig. S2b presents a simulated spectrum measurement observed from the view of frequency tuning, where four red indicating lines i~iv mark the key points related to the spectral evolution process simulated in Fig S2a, providing rich physical information. The intervals between lines i~ii and iii~iv represent the oscillation of dual longitudinal modes, while the interval between lines ii~iii corresponds to the SLM oscillation region. The distance between lines i~iii is equal to the spacing between neighboring longitudinal modes in the cavity (i.e. free spectral range, FSR). The spacing between lines i~iv represents a complete evolution cycle of a longitudinal mode from birth to death (i.e., the output bandwidth of a laser, $\Delta v_{out}$), which directly reflects the total gain linewidth of Ne atoms. One of the experimental results shown in Fig. S2c displays how longitudinal modes and laser intensity evolve with frequency tuning under the condition of He-Ne mixed gas with 9 torr pressure. For a clearer presentation of experimental data, we horizontally arrange the spectral data (based on a higher sampling frequency) on the axis of frequency tuning, compressing the three-dimensional evolution process shown in Fig. S2a, b into a two-dimensional image through temporal panoramic stitching. Comparing Fig. S2b, c, the simulated and experimental results match quite well, which means it is reasonable to consider the range between indicating lines i~iv in Fig. S2c as the output bandwidth in the experiment.

**S3: Peak power of SLM as a function of pressure and temperature**

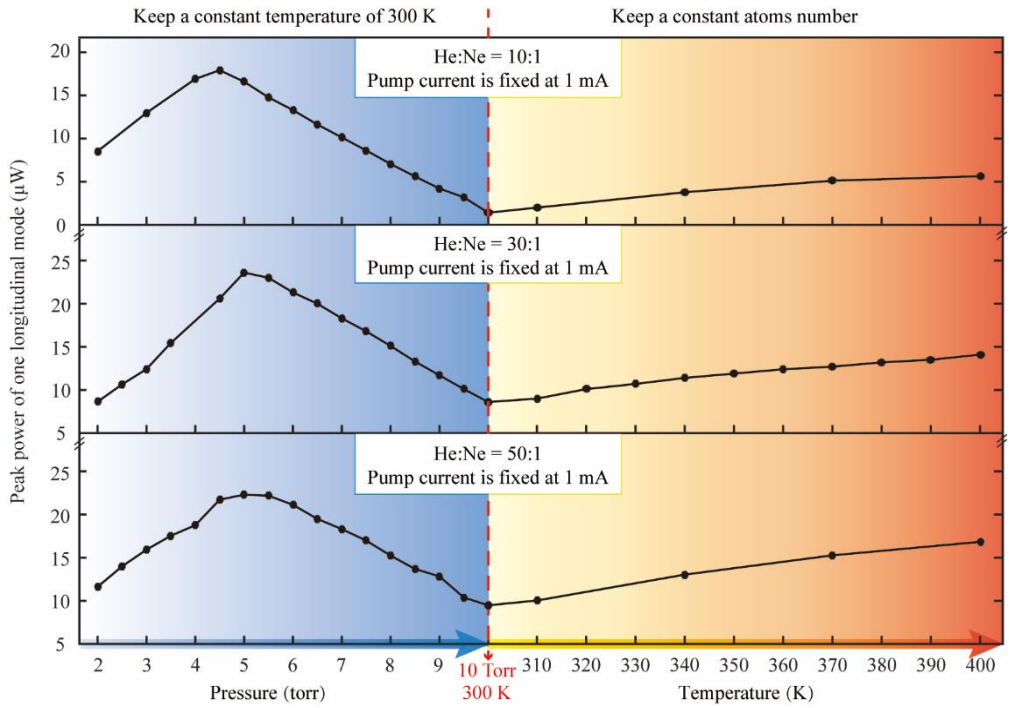

Fig. S3 **Evolution of peak power of one longitudinal mode at a constant pump current of 1 mA.**

As is shown in Fig. S3, regardless of the gas ratio of He and Ne atoms, as long as the pumping current is maintained at a constant level, the peak power of a single longitudinal mode will always initially increase with increasing gas pressure and then continue to decrease. It will also gradually increase with increasing temperature. So, in order to maintain a constant peak power of SLM (i.e. maintain a constant maximum pure gain), we need to change the pump current under different pressure or temperature conditions.

**S4: Detailed experimental data of different He-Ne gas ratios**

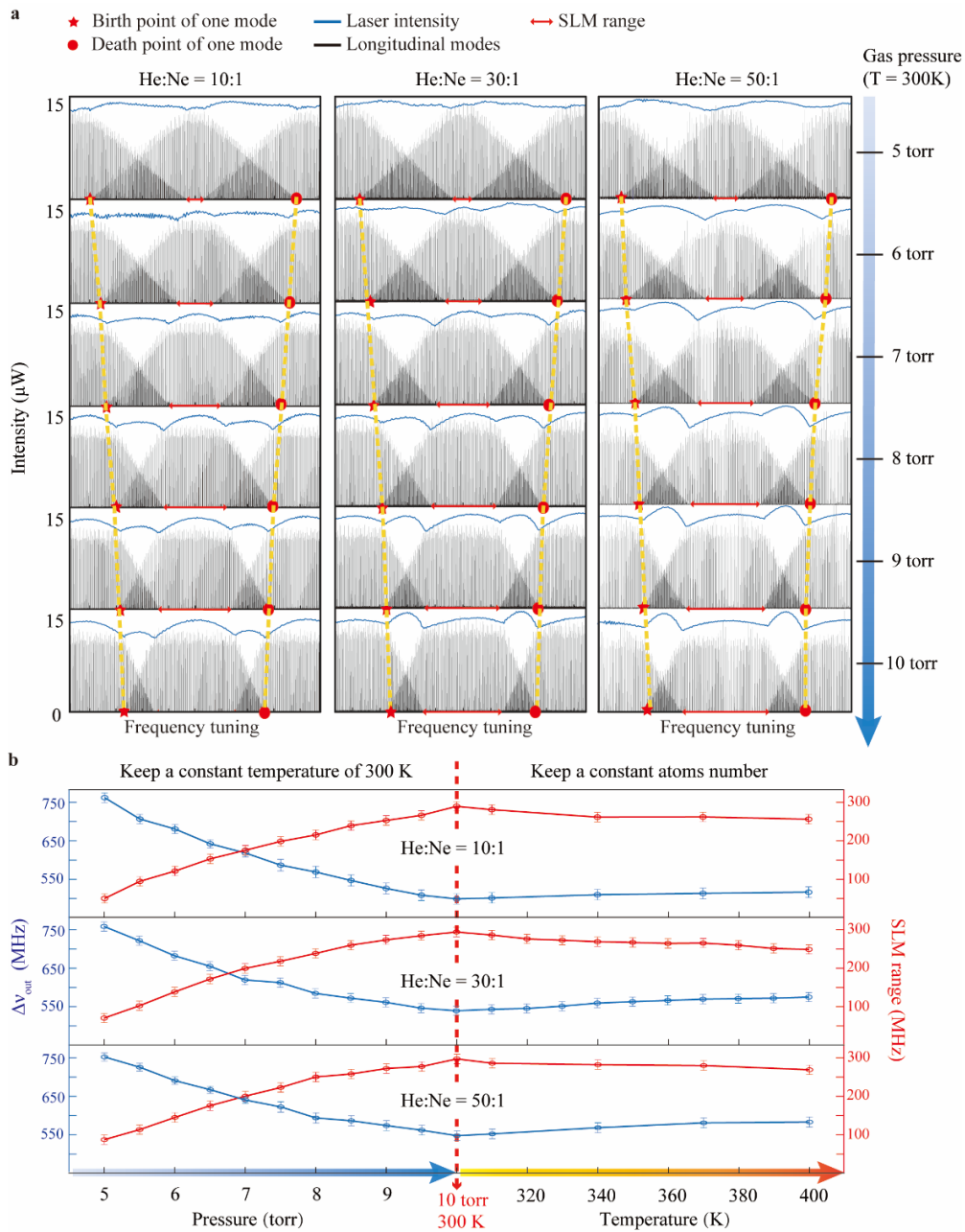

Fig. S4 **Spectral measurements under different He-Ne gas ratios conditions. a** Spectral evolution with increasing pressure at a constant temperature of 300 K. **b** The evolution of output spectral width and SLM range with increasing pressure (temperature is constant) and increasing temperature (total gas atom number is constant).

Fig. S4 indicates the phenomenon of atomic cooling and gain linewidth suppression induced by higher pressure is universal in various He-Ne laser systems.

**S5: Details of experimental setup**

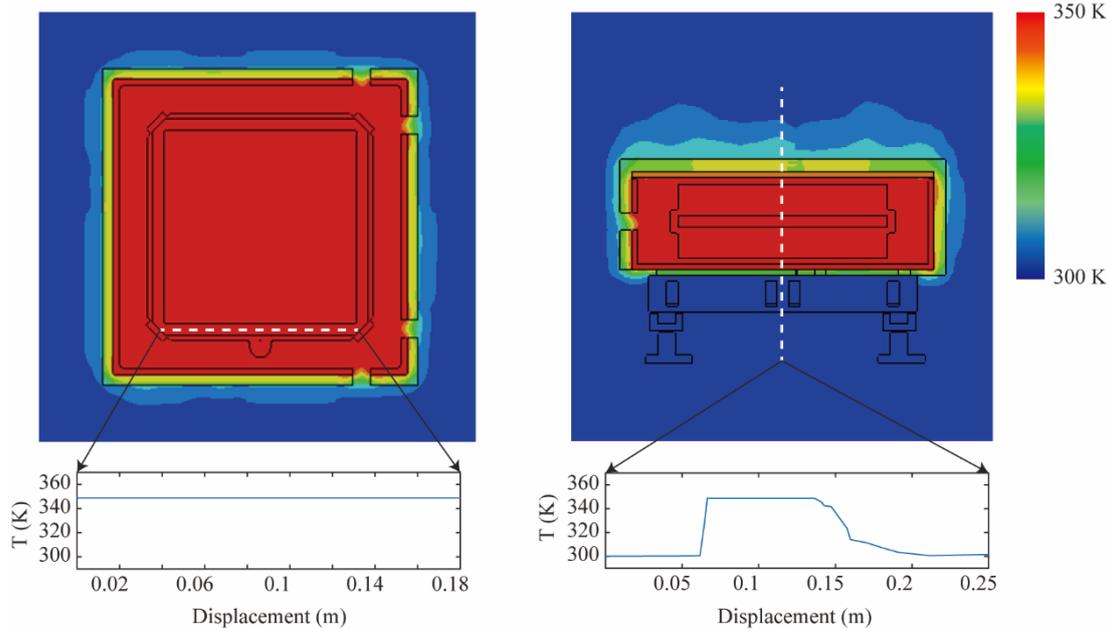

Fig. S5 **Simulation of the heating process at a temperature of 350 K from up (left) and side (right) view.**

As shown in Fig. S5, we designed a thermo-insulation box to ensure the uniformity of heating the laser, and according to the simulation results, it can be seen that the temperature distribution in the gain region (indicated by the white dashed line in the left column) is uniform during the heating process.

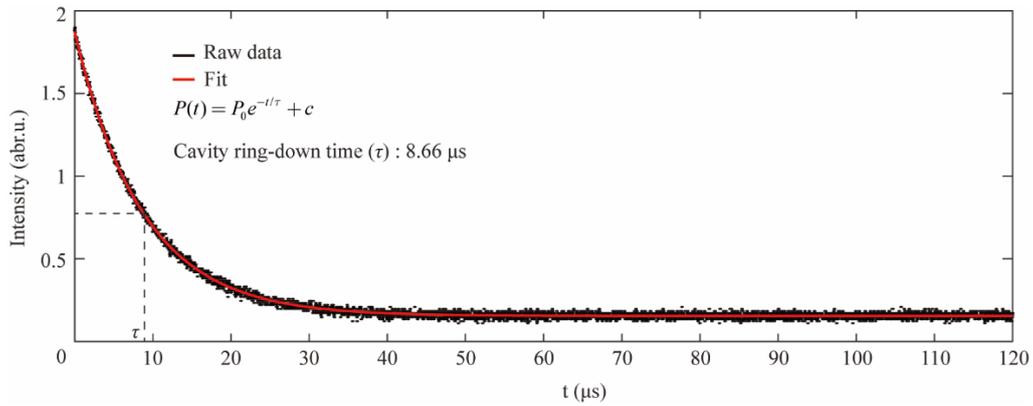

Fig. S6 **Measurement of the total loss of the laser resonator by cavity ring-down technology**

As shown in Fig. S6, the total loss of the laser resonator is measured by cavity ring-down technology, where the raw data is fit by $P(t) = P_0 e^{-t/\tau} + c$ ($P_0$ is the initial output power, $\tau$ is the decay time or cavity ring-down time and $c$ is the background noise.) So the total loss of the laser resonator can be calculated by $\delta = L/(c \times \tau) = 1/(FSR \times \tau)$ =277.3 ppm.